\documentclass[%
 reprint,
nofootinbib,
 amsmath,amssymb,
 aps,
]{revtex4-2}

\usepackage{graphicx}
\usepackage{dcolumn}
\usepackage{enumitem}
\usepackage{bm}

\usepackage{listings}
\usepackage[dvipsnames]{xcolor} 
\usepackage{xspace}
\usepackage{amsmath}
\usepackage{orcidlink}

\usepackage{todonotes}
\let\Oldtodo\todo
\renewcommand{\todo}[1]{\Oldtodo[inline]{\textbf{TODO:} #1}}

\lstdefinelanguage{json}{
  basicstyle=\ttfamily\small,
  numbers=left,
  numberstyle=\tiny\color{gray},
  stepnumber=1,
  numbersep=5pt,
  showstringspaces=false,
  breaklines=true,
  frame=single,
  literate=
   *{0}{{{\color{blue}0}}}{1}
    {1}{{{\color{blue}1}}}{1}
    {2}{{{\color{blue}2}}}{1}
    {3}{{{\color{blue}3}}}{1}
    {4}{{{\color{blue}4}}}{1}
    {5}{{{\color{blue}5}}}{1}
    {6}{{{\color{blue}6}}}{1}
    {7}{{{\color{blue}7}}}{1}
    {8}{{{\color{blue}8}}}{1}
    {9}{{{\color{blue}9}}}{1}
    {:}{{{\color{red}{:}}}}{1}
    {,}{{{\color{red}{,}}}}{1}
    {\{}{{{\color{olive}{\{}}}}{1}
    {\}}{{{\color{olive}{\}}}}}{1}
    {[}{{{\color{olive}{[}}}}{1}
    {]}{{{\color{olive}{]}}}}{1},
}

\lstdefinestyle{json}{
  language=json,
  basicstyle=\ttfamily\small,
  keywordstyle=\color{blue},
  stringstyle=\color{teal},
  commentstyle=\color{gray},
}

\newcommand{\UW}{Department of Astronomy, University of Washington, Seattle, WA, 98195}
\newcommand{\CCA}{Center for Computational Astrophysics, Flatiron Institute, 162 Fifth Ave, New York, NY, 10010, USA}

\newcommand{\site}{\textit{The Software Citation Station}\xspace}
\newcommand{\UCSD}{Department of Astronomy \& Astrophysics, University of California, San Diego, 9500 Gilman Drive, La Jolla, CA 92093, USA}
\newcommand{\UT}{David A. Dunlap Department of Astronomy \& Astrophysics, University of Toronto, 50 St. George Street, Toronto, ON M5S 3H4, Canada}
\newcommand{\DI}{Dunlap Institute of Astronomy \& Astrophysics, University of Toronto, 50 St. George Street, Toronto, ON M5S 3H4, Canada}



\begin{document}

\author{Tom Wagg\,\orcidlink{0000-0001-6147-5761}$^1$}
\email{tomjwagg@gmail.com}
\affiliation{\CCA}
\affiliation{\UW}
  
\author{Floor S. Broekgaarden\,\orcidlink{0000-0002-4421-4962}$^1$}
\email{fbroekgaarden@ucsd.edu} 
\thanks{$^1$ Both Tom and Floor contributed equally to the work.}
\affiliation{\UCSD}

\author{Phil R. Van-Lane\,\orcidlink{0009-0009-4567-9946}}
\email{phil.vanlane@mail.utoronto.ca}
\affiliation{\UT}
\affiliation{\DI}
\affiliation{\UCSD}

\title{Streamlining and standardizing software citations with \\ \site}

\date{\today}

\begin{abstract}
    Software is crucial for the advancement of science, especially in the context of rapidly growing datasets that increasingly require algorithm and pipeline development to process data and produce results. However, software has not always been consistently cited, despite the importance of citations in strengthening support for software development. To encourage, streamline, and  standardize the process of citing software in academic works, we introduce \href{https://flatironinstitute.github.io/software-citation-station/}{`\site'}: a publicly available website and tool to quickly find correct software citation statements and add them to publications. 
\end{abstract}

\maketitle
\setcounter{footnote}{3}


\section{Introduction}

Software is becoming increasingly important as the basis of many scientific results \citep{Hannay:2009}. In particular, software development has become an essential part of every sub-field of astronomy, but is often not adequately supported \citep{decedal:2020}.
Citations are an important tool to properly credit earlier work, value authorship, motivate funding, enable peer-review, validate and reproduce findings, support collaboration and community, and encourage reuse and building on the work of others \citep{katz2020recognizing}. 
While there exists a well-developed ecosystem of tools and services to assist with citations to traditional publications, such as \href{https://ui.adsabs.harvard.edu/about/}{NASA/ADS} and \href{https://arxiv.org/}{arXiv}, this infrastructure does not work as well for software citation.
As a result, software is often not cited properly.
A major issue is that software lacks a standardized citation practice, leaving it to the software author to point readers to the citation mechanism \citep{howison2016software,  Niemeyer:2016, Kai:2017-Rcited, Bouquin:2020-apjs,  Alsudais:2021, Bouquin:2023}.

Some advancements have been made in recent years to promote and standardize software citations in academic journals. These advancements include the creation of specific recommendations for software citation such as the reports by the Joint Declaration of Data Citation Principles, and the recommendations of the FORCE11 Software Citation Working Group \citep{martone2014data, smith2016software, katz2020recognizing}.
Several journals have developed software policies and include software sections or software acknowledgments in their publication templates \citep[e.g.,][]{2016AJ....151...21V, 2019ASPC..523..693T}. Examples include those by \href{https://journals.aas.org/policy-statement-on-software/}{AAS publishing} \citep{AAS_software_policy}, and the examples listed on \href{https://www.chorusaccess.org/resources/software-citation-policies-index/}{CHORUS} \citep{CHORUS_software_policy}. This effort has helped, as evidenced by IRAF \citep[][]{IRAF...1993ASPC...52..173T,IRAF...1986SPIE..627..733T} citations increasing by ${\sim}150\%$ after AAS started actively recommending authors to include their citations in submitted papers \citep{iraf_increases}. AAS Journals now also recommend \site to authors for finding the preferred citations of the software used in their work \citep{AAS_software_policy}.

Other advancements focus on the review and reproducibility of software accompanying research. An important example is the creation of The Journal of Open Source Software (\href{https://joss.theoj.org}{JOSS}\footnote{\protect\url{https://joss.theoj.org}}; \citet{Smith2018JOSS}), an open access journal for research software packages that enables peer-review of software without processing charges or subscription fees. Recently, AAS publishing started a public collaboration with JOSS where authors submitting to one of the AAS journals can also publish a companion software paper in JOSS \citep{vishniac2018cooperative}. Additionally, AAS publishing started to facilitate `living articles' (articles that are formally reviewed but that can be updated to accommodate changing conditions, upgrades, or new functionality) to make it easier to publish and update papers describing (long-lived) software \citep{vishniac2018living}.
Similarly, \href{https://show-your.work/en/latest/}{\textit{Show your Work!}} \citep{Luger2021} provides an automated workflow that creates a self-contained recipe for readers to reproduce paper results based on accompanying code and software.
In addition, \href{https://figshare.com/}{\textit{Figshare}} is a website designed to increase the visibility of software accompanying (arXiv) papers.

Another important focus has been on creating unique identifiers to ensure citations are standardized and accurately counted. The Astrophysics Source Code Library (\href{https://ascl.net/}{ASCL}) is a website and tool designed with this goal, specifically dedicated to curating and indexing software used in astronomy literature by providing unique identifiers. 
In a similar vein, GitHub has created an integration with Zenodo to make it easier to reference and cite GitHub repositories in academic literature \citep{github_software_policy}. 
Software records in places such as ASCL, Zenodo, and JOSS, which provide the software with a unique identifier, will typically also be included in ADS (with citations counted based on the unique identifier), and efforts are underway to streamline this between software versions and platforms, e.g., through the AAS \href{https://journals.aas.org/news/asclepias_aas233/}{\textit{Asclepias}} project \citep[][]{nielsen_2018_1283381, vandeSandt:2019, Henneken2022Asclepias}.

Finally, several Python packages, including \texttt{makecite} \citep{adrian_price_whelan_2019_3533303}, {\texttt{citepy}} \citep{citepy_software}, and \texttt{duecredit} \citep{duecredit}, have been developed to help researchers streamline the process of gathering citations based on the packages imported in a user-provided Python script. However, these packages have some limitations given that (a) they only support Python based software, (b) they can be difficult to apply to projects which use multiple Python scripts, (c) they require installing and running the code, and (d) they still rely on (incomplete) libraries of software references.

Despite these advancements, many authors still do not include proper software citations. This is because it can be challenging to find the citation for many tools, users might not be aware that certain software has references, and/or users might use different keywords or references. This makes it difficult to identify and quantify the impact of software, and to give proper credit to the appropriate software contributors \citep[see][and references therein]{katz:2018software, Katz:2019, katz2020recognizing, vandeSandt:2019, Bouquin:2020-apjs,Bouquin:2023,  Druskat:2021, Druskat:2024}. 
As citing software is becoming increasingly important to support careers and science, it is important that we increase the practice of software citation in academic work. To this end we have created `\href{https://flatironinstitute.github.io/software-citation-station/}{\site}'\footnote{\protect\url{https://flatironinstitute.github.io/software-citation-station/}}, a publicly available website and tool to easily find correct software citations and add them to academic work. \site is publicly available through \citet{Wagg:2024-scs-paper}.


\section{\site}

\begin{figure}
    \centering
    \includegraphics[width=\columnwidth]{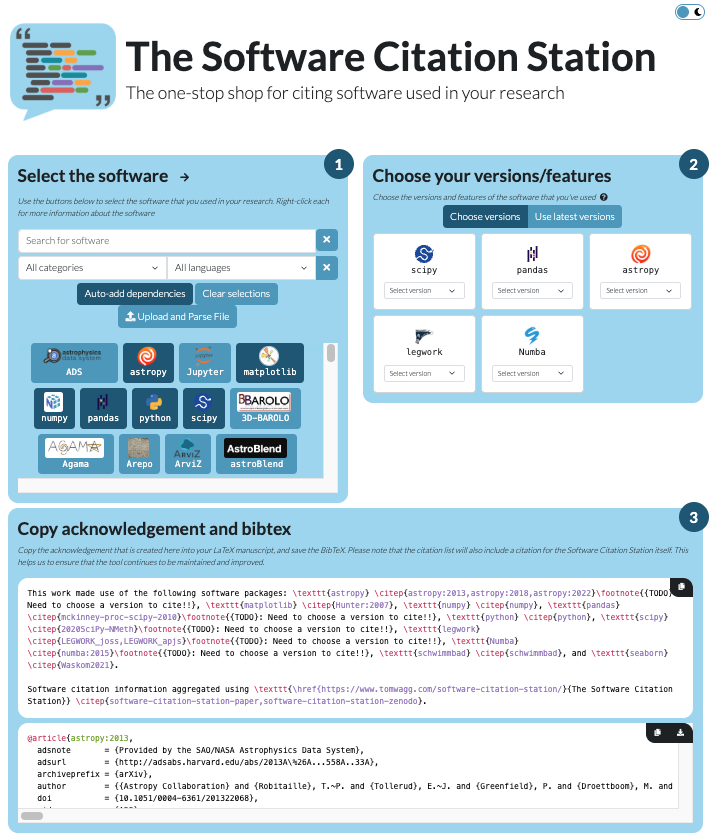}
    \caption{Screenshot demo of using \site to cite software. Panel 1 allows users to search and filter software packages in order to select those used in their research. Panel 2 lets the user select the specific version of the selected software packages (for those that are indexed on Zenodo) and specific features used in their research. Panel 3 is then populated with an acknowledgment for a \LaTeX{} manuscript and the associated BibTeX respectively. Places where the author has to add additional information (e.g., which version of a package was used) are flagged as a TODO footnote in the output.}
    \label{fig:scs-screenshot}
\end{figure}

The primary goal of \site is to streamline and standardize the process of citing software in academic work. We streamline the process by designing the site's core functionality to be a simple three-step process (see Figure~\ref{fig:scs-screenshot}). Citations are standardized by adhering to the requirements outlined by each software package.

\subsection{Core functionality}

Users first select the software that they have used in their academic work in panel 1 (see Figure~\ref{fig:scs-screenshot}). One can search for particular keywords or package names, as well as filter by category or programming language to refine the list. Users can right-click each item to see further details about the package such as a short description, a link to the documentation and the associated Zenodo record.
Users can also choose whether or not to automatically add the dependencies of each package (which are stored by the website) to be included in the citation statement. Finally, there is an option to upload a \texttt{requirements.txt} or \texttt{environment.yml} file instead of manually adding each software package, which our tool parses and extracts the packages from.

As each package is selected, if it is indexed on Zenodo, \site queries the Zenodo API for the latest list of available versions of the software and populates a dropdown with the options. Users can then select the version used in their research for each package in panel 2.
Since some software has different citation recommendations for different sub-packages or features, our tool also prompts users to identify the specific features used (where applicable) at this step.

As this selection occurs, an acknowledgment and associated BibTeX is generated in panel 3. Users can directly copy the acknowledgment into their \LaTeX{} manuscript, either in the acknowledgments section or, for AAS journals, in the \texttt{Software} section. The BibTeX can either be downloaded directly as a BibTeX file or copied to the clipboard to insert into another file.

\subsection{Custom citation statements}\label{sec:custom}
There is not a universally accepted software citation standard, so in many cases software packages have differing requirements for how they are cited. Some software authors request that users write a specifically worded acknowledgment in their work, in addition to particular citations. We allow for flexibility in how a software package is cited, and arbitrary customization is possible when adding a new package to \site (see Section~\ref{sec:new_software}).

We note that this custom citation statement has a clear drawback, in that it works contrary to our goal of providing a standardized interface for software citation. During the moderation phase of submitted software we encourage authors to use a more standard citation format to avoid this issue. However, in some cases, changing the citation format is not possible and so we allow for flexibility in these cases.

\subsection{Package dependencies}\label{sec:deps}
It is very common for open-source projects to depend on other software packages, which can often be integral to the results produced by the top-level software. It is important to consistently cite these dependencies of software \citep{sochat2022citelang}. By default, each time a package is selected on \site, its dependencies (and dependencies of \textit{those} dependencies) that are not already selected will be added to the list of packages to cite. The user is notified of these additional selections.

However, in some cases, adding the full list of dependencies may not be an appropriate choice for a paper. With this in mind, users can toggle a button for whether to automatically add dependencies of selected software. This option allows users to add only the specific dependencies that they deem necessary for citation.

\subsection{Adding new software}\label{sec:new_software}
\begin{figure}
    \centering
    \includegraphics[width=\columnwidth]{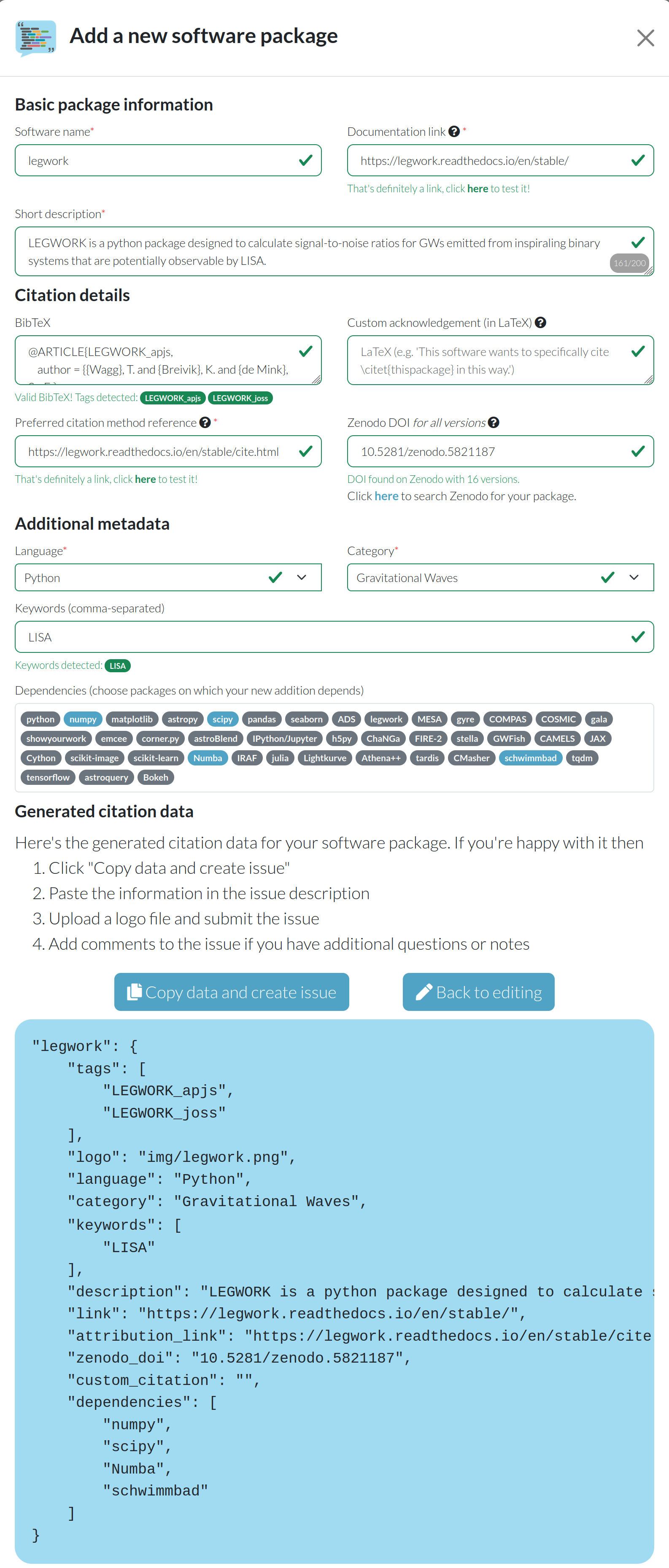}
    \caption{Screenshot of \href{https://flatironinstitute.github.io/software-citation-station/?new-software=true}{the online form} to help users add new software to \site. \texttt{LEGWORK} \citep{LEGWORK_joss,LEGWORK_apjs} is used as an example. The data generated using this form can directly be uploaded as a GitHub issue.}
    \label{fig:scs-new-software-screenshot}
\end{figure}

We aim for \site to be open to all software packages used in academic work. As such, the site includes \href{https://flatironinstitute.github.io/software-citation-station/?new-software=true}{a form for submitting new software packages}, which guides users through the submission process. Figure~\ref{fig:scs-new-software-screenshot} shows an example of the online form. Each package requires the following metadata:
\begin{itemize}[itemsep=0pt]
    \item \texttt{tags}: a list of the BibTeX tags associated with this package.
    \item \texttt{language}: the programming language of the software.
    \item \texttt{category}: the category of the package (e.g., ``population synthesis'' or ``visualization'').
    \item \texttt{keywords}: a list of relevant keywords for searching.
    \item \texttt{description}: a short 1--2 sentence description of the package.
    \item \texttt{link}: a URL to the documentation, or code repository if no documentation is available.
    \item \texttt{attribution\_link}: a link to a website that outlines the preferred citation format of the package.
\end{itemize}
as well as the following optional fields:

\begin{itemize}[itemsep=0pt]
    \item \texttt{zenodo\_doi}: the DOI associated with all versions of the package on Zenodo.
    \item \texttt{custom\_citation}: a custom citation string that overwrites the default generated citation (see Section~\ref{sec:custom} for more details).
    \item \texttt{dependencies}: a list of package names that should also be cited if this package is cited (see Section~\ref{sec:deps}).
    \item \texttt{feature\_tags}: a list of any specific features of the software that have different associated citations.
\end{itemize}
The form is designed to streamline the submission of new packages and catch simple errors automatically. \texttt{tags} are extracted directly from submitted BibTeX and languages/categories can be selected from a dropdown menu populated based on currently known packages. Additionally, prior to submission, the form is validated in several ways such as requiring URLs to direct to existing websites. The form also checks that the Zenodo DOI corresponds to an existing record with multiple versions, and if a user has submitted the DOI that only corresponds to a specific version, it will correct it. The form also provides the option to upload a logo for the software.

Once the user has filled out the form, it is a one-click process for them to generate a GitHub issue with the required metadata auto-populated in JSON format. This is then copied directly into the website code, at which point the new software appears immediately in \site.

\subsection{Community maintenance}\label{sec:maintenance}

\site is maintained by community volunteers who review every submission before it is added to the website. Since new packages are submitted through GitHub, each submission and its associated discussion takes place in a public issue or pull request, which any member of the community can view and comment on. This allows both software authors and users to flag potential problems before an entry is added.

During review, each submission is checked against the citation instructions provided in the online documentation of the corresponding package (linked via the \texttt{attribution\_link} field, see Section~\ref{sec:new_software}). This ensures that the citations generated by \site follow the requirements set out by the software authors, rather than adding inconsistent or incorrect references to the literature. This review is also the stage at which we encourage authors to adopt a standard citation format where possible (see Section~\ref{sec:custom}).

Software and its citation requirements naturally change over time, for example when a new paper describing a package is published or its Zenodo record is updated. Any outdated or incorrect entry can be corrected by opening a new issue with the ``Update software'' template, which goes through the same public review process as a new submission. Additionally, since the list of available versions for packages indexed on Zenodo is queried directly from the Zenodo API, these entries automatically remain up to date as new versions are released.


\section{Future work}

We created \site to make it incredibly easy for researchers to add citations of commonly-used software to their academic publications. This is in line with recommendations by \citet{Bouquin:2023} who mentioned the need to ``Create a website that can act as a central source of information on software citation in practice''.\footnote{See also \href{https://citeSoftware.org/}{CiteSoftware.org}, \href{https://citeas.org/}{CiteAs.org}, and \href{https://cite.research-software.org/}{cite.research-software.org} which are online guides and tools on how to cite software, but which do not yet create the BibTeX for a specific software package for you as \site does.}
In the longer term it would be beneficial to integrate \site with other existing efforts to improve citations of software. 
This includes connecting or integrating our website as part of arXiv, ADS, \href{https://scixplorer.org/}{SciX}, and/or ASCL, as well as connecting to citation software like \texttt{makecite} \citep{adrian_price_whelan_2019_3533303}, {\texttt{citepy}} \citep{citepy_software}, and \texttt{duecredit} \citep{duecredit}. 
Specifically, it would be useful to create a persistent library with citation entries that is community-updated to streamline consistency and completeness of different software citation tools. Ideally such a library could be generated from data provided by the software developer.  
An important step to achieve this will be for software developers in academia to use consistent citation practices, which could include adding a citation format file (e.g.,\ \href{https://citation-file-format.github.io/}{\texttt{CITATION.cff}}, \citet{citation-cff}) and/or a CodeMeta file (\href{https://codemeta.github.io/}{CodeMeta.json}, \citet{jones2017codemeta}) to GitHub or other software repositories, and assigning software references a unique identifier (see \citet{Katz:2014, Cosmo:2020}, and references therein). Additionally, such software citations should also support version control \citep[e.g.,][]{Chen:2020}. Such work is out of scope of this paper, but there are examples of recent efforts in this direction such as the HERMES  \citep{Druskat:2022}, CiteAs \citep{CiteAs2021}, and Asclepias \citep[][]{Henneken2022Asclepias} projects.
Further work in defining and encouraging these best practices will be an important topic to discuss and pursue in the future \citep[cf.][]{katz:2018software, Katz:2019, katz2020recognizing, vandeSandt:2019, Cosmo:2020, 2021arXiv211114278A, Druskat:2022, Bouquin:2023}. As a step in this direction, the 2026 Software Citation Workshop brought together researchers, software developers, librarians, and publishers to discuss these challenges, resulting in a practical primer on software citation for researchers \citep{Broekgaarden+2026:2026arXiv260928622B} and a set of community best practices \citep{Van-Lane+2026:2026arXiv261000528V}.


\section{Additional resources}
Several resources for best software citation practices have been created. In particular, The FORCE11 Software Citation Implementation Working Group (\href{https://github.com/force11/force11-sciwg}{SCIWG}\footnote{\url{https://github.com/force11/force11-sciwg}}), which has been dedicated to implementing the software citation principles from the FORCE11 Software Citation Working Group, provides an excellent collection of resources.

Multiple groups have written software citation checklists for authors \citep{smith2016software, chue_hong_2019_3479199,  katz2020recognizing}, developers \citep{chue_hong_2019_3482769, albert_2019_3581326, 2020arXiv201213117B}\footnote{See also the guidelines by \href{https://docs.github.com/en/repositories/archiving-a-github-repository/referencing-and-citing-content}{GitHub}, \href{https://ascl.net/home/getwp/351}{ASCL}, \href{https://www.astrobetter.com/blog/2019/07/01/citing-astronomy-software-inline-text-examples/}{astrobetter}, and \href{https://cfa-library.github.io/citesoftware.org/}{cite-software}.}, and journals \citep{katz2020recognizing, stall2023journal}. A summary is also provided by \citet{Katz:2019} in section 3.5 (how to cite software in text) and sections 3.2 and 3.3 (metadata for software citation). \citet{puebla2024ten} provide a guide to recognizing data and software contributions in hiring, promotion, and tenure in academia.  \citet{Wofford:2020, erdmann_2021_5651648}, and \citet{AAS_jupyter_notebook_policy} provide recommendations for making Jupyter notebooks reproducible and citeable. We also direct the reader to the \href{https://cfa-library.github.io/citesoftware.org/resources/}{CiteSoftware.org} webpage which has resources related to software citation, credit, and preservation.

Several organizations are leading software and coding focused academic schools, many of which also include publicly available recordings and lecture notes. Examples include \href{https://github.com/semaphoreP/codeastro}{{Code/Astro}}, the \href{https://lsstdiscoveryalliance.org/programs/data-science-fellowship/}{Legacy Survey of Space and Time Corporation (LSSTC) data science fellowship program}, and the \href{http://lssds.aura-astronomy.org/winter_school/}{La Serena School for data science}. In addition, the Astronomical Data Analysis Software and Systems (\href{https://www.adass.org}{ADASS}) organizes annual software-focused conferences and workshops in astronomy. 

Other software-focused developments in astronomy include  several organizations initiating software development prizes to recognize and highlight software contributions in the field, such as the \href{https://www.adass.org/softwareprize.html}{ADASS Prize for an Outstanding Contribution to Astronomical Software} and the \href{https://www.nasa.gov/ames-exploration-technology/awards/}{NASA Software of the Year Award}. Additionally, there are efforts to promote more stable career opportunities for software focused astronomers including the \href{https://www.simonsfoundation.org/grant/scientific-software-research-faculty-award/}{Software Research Faculty Award} from the Simons Foundation. These efforts are important to address the issues beyond citing software, such as supporting and funding the careers of software developers and software engineers in astronomy \citep[see the discussions in][]{2020arXiv200501469A, Bouquin:2023, carlin2023all, sochat2024infrastructure}. Moreover, the development of tools such as \textsc{CiteLang} \citep{sochat2022citelang} moves beyond simple software citations to contributions of the dependencies for software. Finally, we point readers to \citet{Smith2022} for an example of a career path as a software engineer in astronomy and a discussion of the challenges.

\section{Data availability and software}
\site is publicly available on Zenodo at \citet{Wagg:2024-scs-paper}. We welcome contributions to the code which can be made through GitHub or the website.

\begin{acknowledgments}
TW acknowledges support from NASA ATP grant 80NSSC24K0768. This work was supported by Simons Foundation award 1141468 hosted at Columbia University. This research has made use of NASA's Astrophysics Data System.  The authors acknowledge support for the 2026 Software Citation Workshop, which was supported by the NASA TWSC program under award number 316634-00001. PvL and FSB acknowledge support from NASA HPOSS grant 80NSSC25K7555 under award number 316592-00001. FSB  acknowledges support from NSF 22-624 Astronomy and Astrophysics Research Grants under award number 2606407. 
\end{acknowledgments}


\bibliography{software-biblio}
\bibliographystyle{apsrev4-2}
\end{document}